\documentclass[twocolumn,showpacs,preprintnumbers,amsmath,amssymb,prl,superscriptaddress]{revtex4-1}

\usepackage{graphicx}
\usepackage{color}
\usepackage[colorlinks=true,plainpages=false,linkcolor=blue,urlcolor=blue,citecolor=blue,pdfpagemode=UseNone,pdfstartview=FitBH]{hyperref}

\def\SIO{$\rm Sr_2IrO_4$}
\def\bSIO{$\rm Sr_3Ir_2O_7$}
\def\jeff{$j_{\rm{eff}}$}

\begin{document}

\title{Two-magnon Raman scattering and pseudospin-lattice interactions in $\bf Sr_2IrO_4$ and $\bf Sr_3Ir_2O_7$}
\author{H.~Gretarsson}
\author{N. H.~ Sung}
\author{M. H\"oppner}
\author{B. J.~ Kim}
\author{B.~ Keimer}
\author{M. Le Tacon}
\affiliation{Max-Planck-Institut f\"{u}r Festk\"{o}rperforschung, Heisenbergstr. 1, D-70569 Stuttgart, Germany}

\date{\today}

\begin{abstract}
We have used Raman scattering to investigate the magnetic excitations and lattice dynamics in the prototypical spin-orbit Mott insulators \SIO\ and \bSIO.
Both compounds exhibit pronounced two-magnon Raman scattering features with different energies, lineshapes, and temperature dependencies, which in part reflect the different influence of long-range frustrating exchange interactions. Additionally, we find strong Fano asymmetries in the lineshapes of low-energy phonon modes in both compounds, which disappear upon cooling below the antiferromagnetic ordering temperatures. These unusual phonon anomalies indicate that the spin-orbit coupling in Mott-insulating iridates is not sufficiently strong to quench the orbital dynamics in the paramagnetic state.
\end{abstract}


\maketitle

\noindent

The recent discovery of Mott-insulating states driven by the confluence of intra-atomic spin-orbit coupling and electronic correlations (``spin-orbit Mott insulators'') has triggered a wave of research on novel magnetic ground states and excitations. The most widely studied spin-orbit Mott insulators are iridium oxides based on Ir$^{4+}$ ions with electron configuration $5d^5$ and total angular momentum \jeff $= \frac{1}{2}$. Mott-insulating iridates with \jeff-pseudospins arranged on geometrically frustrated lattices are promising candidates for spin-liquid states with unusual transport properties and fractionalized excitations. \cite{Okamoto_2007,Jackeli_PRL2009,Chaloupka_PRL2010,Shitade_PRL2009,review} Iridates with square-lattice geometries, on the other hand, have been extensively investigated as possible analogues of the cuprate high-temperature superconductors whose low-energy excitations are characterized by the pure spin quantum number $S = \frac{1}{2}$. Indeed, resonant inelastic x-ray scattering (RIXS) experiments ~\cite{Sr2IrO4-RIXS} have shown that the magnon excitations of the prototypical square-lattice antiferromagnet \SIO~ are well described by the Heisenberg model, in close analogy to those of the isostructural Mott-insulator La$_2$CuO$_4$. Very recently, experiments on doped \SIO~ have uncovered tantalizing evidence of a Fermi surface split up into disjointed segments (``Fermi arcs'') \cite{Kim_Science2014} and a low-temperature gap with $d$-wave symmetry \cite{Kim_dWave,Feng_dWave}, which are hallmarks of the doped cuprates. \cite{Keimer_Nature}

This rapidly advancing research frontier has raised fundamental questions about the microscopic electronic structure of spin-orbit Mott insulators. A particularly important question concerns the orbital degeneracy, which plays a central role in the phase behavior of $3d$-electron materials, where the spin-orbit coupling is negligible. In the widely studied manganates, for instance, the orbital degeneracy of the manganese ions is only partially lifted by crystalline electric fields, and coupling of low-lying orbital excitations to lattice distortions leads to the formation of polarons which dominate the physical properties of the doped manganates. In Mott-insulating cuprates, on the other hand, the $S = \frac{1}{2}$ moments of the orbitally non-degenerate Cu$^{2+}$ ions do not couple significantly to the crystal lattice, so that spin fluctuations and their coupling to charge carriers play the major role in most theoretical descriptions of the doped cuprates. \cite{Keimer_Nature} In the Mott-insulating iridates, the orbital degeneracy is lifted by the strong spin-orbit coupling, and the spatial isotropy of the \jeff $= \frac{1}{2}$ ground state of the spin-orbit Hamiltonian is believed to be responsible for the isotropic Heisenberg exchange interactions in undoped \SIO. \cite{Jackeli_PRL2009} The spin-pseudospin correspondence has also been invoked to explain the analogous physical properties of doped iridates and cuprates. \cite{FaWang_PRL2011,Yang_PRB2014} However, the \jeff $= \frac{3}{2}$ excited state in the iridates is found at much lower energy than crystal-field excitations in the cuprates. \cite{Kim2014} Depending on the relative strengths of spin-orbit and crystal-field interactions, static or dynamic lattice distortions can therefore induce significant \jeff $= \frac{3}{2}$ admixtures into the \jeff $= \frac{1}{2}$ ground state, possibly resulting in electron-phonon interactions akin to the manganates. Recent experiments on metallic iridates have indeed uncovered evidence of strong electron-phonon interactions \cite{Proepper2014}, but their origin and their relation to the electronic level hierarchy in Mott-insulating iridates remain unclear.

Raman scattering is well suited as a probe of spin-orbital-lattice interactions in the iridates, because both magnons and phonons can be detected with high resolution and signal intensity. To this end, we have performed Raman scattering experiments on two prototypical Mott-insulating iridates with square-lattice networks of iridium ions, \SIO\ and \bSIO. Both compounds exhibit different Mott gaps \cite{Moon_TdepOptics:PRB2009,Park_TdepOptics:PRB2014} and magnetic ground states \cite{BJKim2009,JWKim_RMXS327:PRL2012,Sr2IrO4-RIXS,Sr3Ir2O7-RIXS,Moretti_Sr327}. In contrast to previous work, \cite{Sr214_Raman,Sr214_Ru_Raman} we observe well defined two-magnon features in the Raman spectra of both systems. The two-magnon excitations in \SIO\ remain visible well above the N\'eel temperature $T_N$, as observed in the insulating cuprates. In \bSIO, on the other hand, they are rapidly suppressed upon heating and are no longer detectable at $T_N$, suggesting strong damping by another set of excitations. In both compounds, we also observe pronounced Fano lineshapes of selected phonons for $T > T_N$, reflecting strong coupling to a low-energy excitation continuum. The marked self-energy renormalization of these phonons upon cooling below $T_N$ demonstrates that this continuum arises from low-energy fluctuations of the \jeff-pseudospins. The discovery of strong and unusual pseudospin-lattice interactions indicates an unquenched orbital dynamics that must be considered in the theoretical description of insulating and metallic iridates and other $5d$-electron materials.

\begin{figure}[htb]
\includegraphics[trim=0.0cm 0cm 0.0cm 0cm, width=0.9\columnwidth]{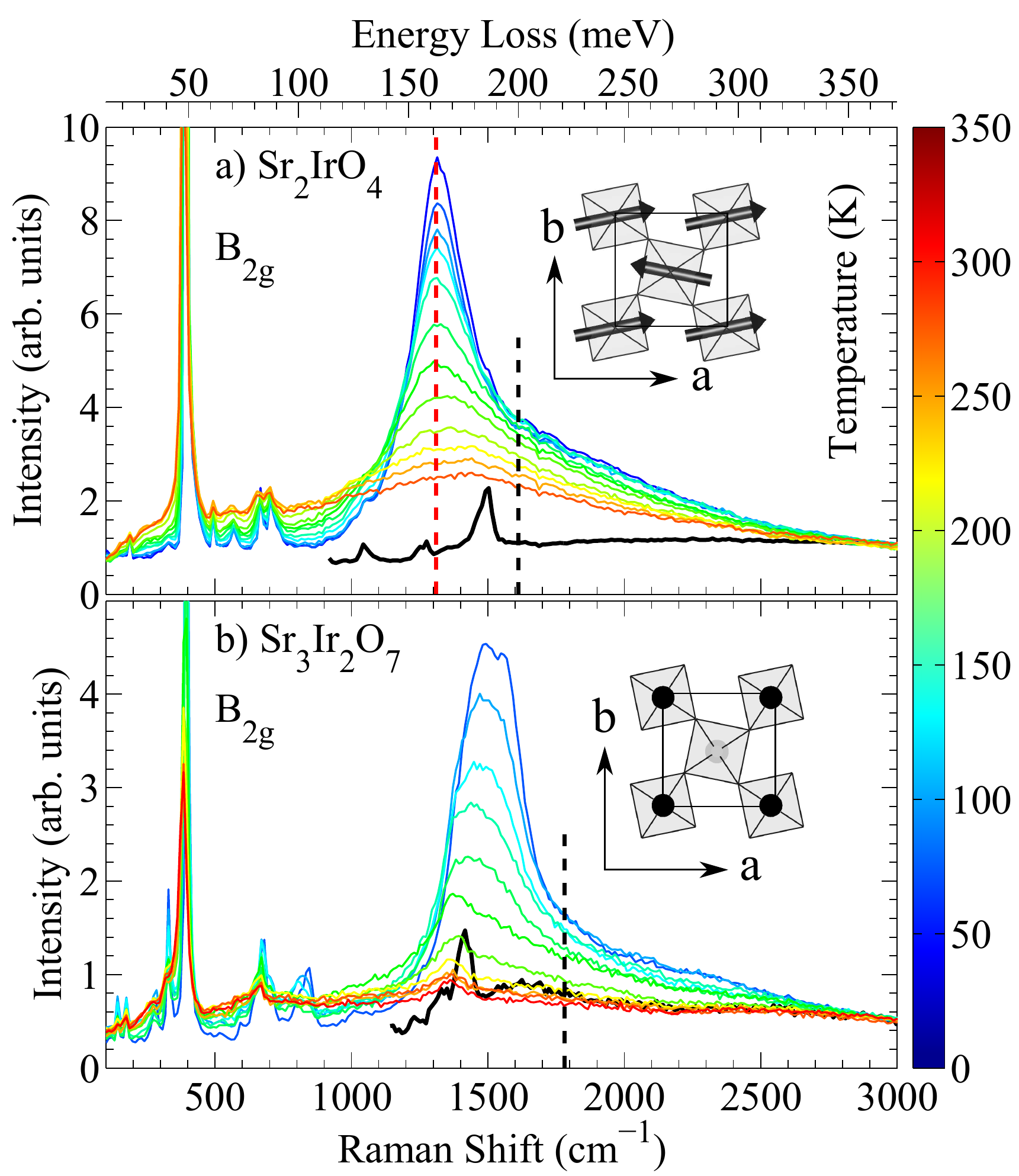}
\caption{\label{RawTwoMagnon} Raman spectra  of the $B_{2g}$ modes at 10 K$<$T$<$350 K for (a) \SIO\ and (b) \bSIO.  A two-magnon scattering becomes visible upon cooling. The thick black lines in both panels show the high-energy $A_{1g}+B_{1g}$ Raman modes at the lowest temperature. For clarity the low energy phonon structure has been removed. The dashed vertical lines indicate the energy of the two-magnon according to the broken bond model (see text) while the red dashed line in (a) represents the energy of the two-magnon according to nearest-neighbor model. Inset shows the $ab$-plane of the systems, arrows indicate magnetic order.}
\end{figure}

The Raman scattering experiments were performed using a JobinYvon LabRam HR800 spectrometer and the 632.8 nm excitation line of a HeNe laser. Resonance effects were checked using different wavelengths \cite{sup}. Two setups were used, low-resolution (600 grooves/mm grating yielding energy resolution $\sim$ 5 cm$^{-1}$) and high-resolution (1800 grooves/mm  $\sim$ 1.5 cm$^{-1}$). Due to the weak intensity of the lowest-energy phonon mode in \SIO\ the  power of the laser was kept at 3.5 mW, for all other measurements a power of 1.5 mW was used. The diameter of the beam was $\sim 10$ $\mu$m. All spectra were corrected for heating and the Bose factor \cite{sup}. The samples were placed in a He-flow cryostat, and the measurements were carried out in backscattering geometry with the light propagating along the crystalline c-axis, while the polarization of the incident and scattered light was varied within the $ab$-plane. Because of rotation of the IrO$_6$ octrahedra around the c-axis in both Sr$_2$IrO$_4$ (space group $I4_1/acd$ $a=b\approx 5.5 {\rm \AA}$, $c\approx 26  {\rm \AA}$) \cite{Sr214_Structure} and Sr$_3$Ir$_2$O$_7$ (space group $Bbca$ $a\approx b\approx 5.5  {\rm \AA}$ and $c\approx 21  {\rm \AA}$) \cite{Cao_TransportSr327:PRB2002}, a- and b-axes are taken along the Ir-Ir next nearest-neighbor direction (see inset Fig. \ref{RawTwoMagnon}). For simplicity we neglect the small orthorhombic distortion of \bSIO\ and refer to the phonons using the tetragonal notation (D$_{4h}$ point group). In this notation the Raman spectrum in the XX-channel probes the $A_{1g} + B_{1g}$ symmetry modes while the XY-channel probes the $B_{2g}$ symmetry modes. Crystals  were grown using a flux method \cite{Nakheon_CrystalGrowth}. Magnetization measurements give N\'eel temperatures of $T_N=$ 240 K for \SIO\ and $T_N=$ 285 K for \bSIO. To obtain a clean surface for the measurements, the crystals were cleaved $ex situ$.

In Fig. \ref{RawTwoMagnon} we plot the $B_{2g}$ Raman spectra of (a) \SIO\ and (b) \bSIO\ between 100 and 3000 cm$^{-1}$ for 10 K$<$T$<$350 K. As the temperature is decreased a sharp asymmetric feature appears centered around  1300 cm$^{-1}$ ($\sim$ 160 meV) and 1500 cm$^{-1}$ ($\sim$ 185 meV) in \SIO\ and \bSIO, respectively. For comparison we show as thick black lines the $A_{1g}+B_{1g}$  high-energy Raman spectra at the lowest temperature which do not show this feature, confirming the $B_{2g}$ nature of the modes. The high energy and the $B_{2g}$ symmetry of these modes strongly suggests that they originate from a two-magnon scattering process \cite{Sr2IrO4-RIXS,Sr3Ir2O7-RIXS}.

Earlier, Cetin \emph{et al.} found evidence for two-magnon Raman scattering in \SIO\ using the XU-channel (unpolarized scattered light) and a yellow laser ($\lambda=561$ nm)  \cite{Sr214_Raman}. Although the energy of this signal was comparable to ours, it was much weaker and exhibited very little structure, in contrast to our observation. We have confirmed that our two-magnon signal can also be observed using $\lambda=568$ nm, \cite{sup} ruling out different laser energies as the reason for this difference. Additionally, when using $\lambda=488$ nm, we could not see the broad high-energy background reported in Ref. \onlinecite{Sr214_Raman}. This suggests that the difference originates from an off-stoichiometric composition of the sample used by Cetin \emph{et al.}, as pointed out in Ref. \onlinecite{Reznik_PRB2015}.

The high quality of our data allows us to compare the energies and lineshapes of the two-magnon peaks with theoretical predictions. Since the two-magnon density of states is usually dominated by zone-boundary excitations that can be viewed as local spin flips, it is instructive to compare the energy of the peaks with simple broken-bond counting arguments. \cite{Loudon_Fleury}  The magnetic interaction parameters for \SIO\ have been estimated by fitting the magnon dispersions measured by RIXS with a spin-wave model~\cite{Sr2IrO4-RIXS} that includes the nearest-neighbor exchange coupling ($J=60$ meV) as well as next-nearest ($J^{\prime}$=-20 meV) and third-nearest neighbor couplings ($J^{\prime\prime}$=15 meV). This places the two-magnon peak at $3J-4J^{\prime}-4J^{\prime\prime}=200$ meV. Since the frustrating antiferromagnetic coupling $J^{\prime\prime}$ is approximately compensated by the $J^{\prime}$ contribution which is of comparable magnitude, this estimate is completely dominated by the nearest-neighbor contribution. We note that in a 2D Heisenberg $S= \frac{1}{2}$ system with only nearest-neighbor coupling $J$, quantum fluctuations reduce the energy of the two-magnon peak from $3J$ to $\sim 2.7J$ \cite{Weber_TwoMagnon:PRB1989}. This places the two-magnon peak at 162 meV, precisely where the main peak is observed in \SIO. According to recent theoretical work, the high-energy tail might then arise from amplitude fluctuations of the magnetic order parameter ~\cite{Zwerger_Higgs}.

The same argument for \bSIO\ yields an estimate of $3J-4J^{\prime}-4J^{\prime\prime}+J_C-4J_{2C}=221$ meV, where $J$=97.4 meV, $J^{\prime}$=11.9 meV, $J^{\prime\prime}$=14.6 meV, $J_C$=59.5 meV, and $J_{2c}=6.2$ meV \cite{Sr3Ir2O7-RIXS}. In this case, both the next-nearest-neighbor and the third-nearest-neighbor in-plane correlations are frustrating, and their combined contribution to the two-magnon energy is much larger than in \SIO. Consequently, the nearest-neighbor model with quantum corrections mentioned above yields an energy much larger ($2.7J \sim 263$ meV) than that of the experimentally observed peak ($\sim 185$ meV). Whereas the quantitative assessment of the impact of long-range exchange interactions on magnetic Raman scattering in the iridates remains a challenge for future model calculations, we note that our observations are in qualitative agreement with recent theoretical work on iron pnictides that demonstrate a large influence of frustrating exchange interactions on the two-magnon Raman spectrum~\cite{Chen_2Magnons}.

\begin{figure}[htb]
\includegraphics[trim=0.0cm 0cm 0.0cm 0cm, width=1\columnwidth]{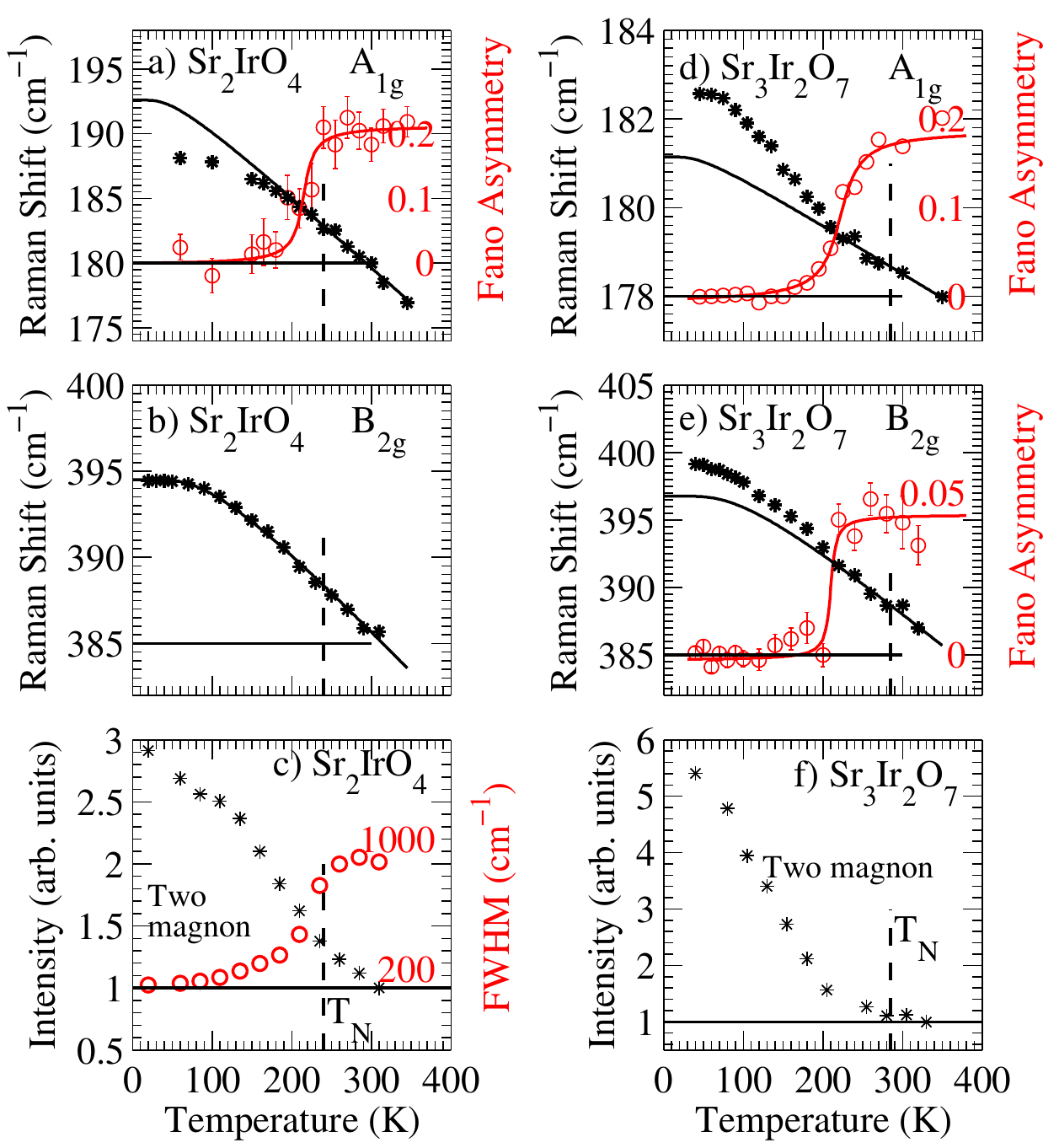}
\caption{\label{IntMagnon} Temperature dependence of the frequency shift (left axis) and Fano asymmetry parameter (right axis) of the (a) $A_{1g}$  and (b) $B_{2g}$ Raman modes of \SIO. (c) Temperature dependence of the two-magnon scattering intensity in \SIO\ compared to its linewidth. The intensity was integrated over a 200 cm$^{-1}$ window centered at the peak position. In (d-f) the equivalent parameters for \bSIO\ can be seen.}
\end{figure}

Figures \ref{IntMagnon} (c) and (f) show the temperature dependence of the energy-integrated intensities of the two-magnon peaks in \SIO\ and \bSIO, respectively, normalized to their value at the highest temperature. In both compounds, the two-magnon peaks broaden and weaken systematically with increasing temperature, as expected from magnon-magnon interactions. In \SIO, the two-magnon peak persists above T$_N$, reflecting the quasi-two-dimensional nature of the spin-spin correlations, as in the case of La$_2$CuO$_4$ \cite{LCO_Raman_TwoMagnon}. In \bSIO, on the other hand, the two-magnon peak intensity is reduced much more strongly upon heating and vanishes around $T_N = 285$ K, in good agreement with RIXS measurements \cite{Sr3Ir2O7-RIXS}. These observations indicate an additional damping channel arising from coupling to a different set of excitations. Potential candidates include charge excitations across the small Mott gap, spin-orbit excitons, \cite{Kim2014} and phonons \cite{Cardona_PRB1990:TwoMagnon}, or combinations thereof.

\begin{figure}[htb]
\includegraphics[trim=0.0cm 0cm 0.0cm 0cm, width=1\columnwidth]{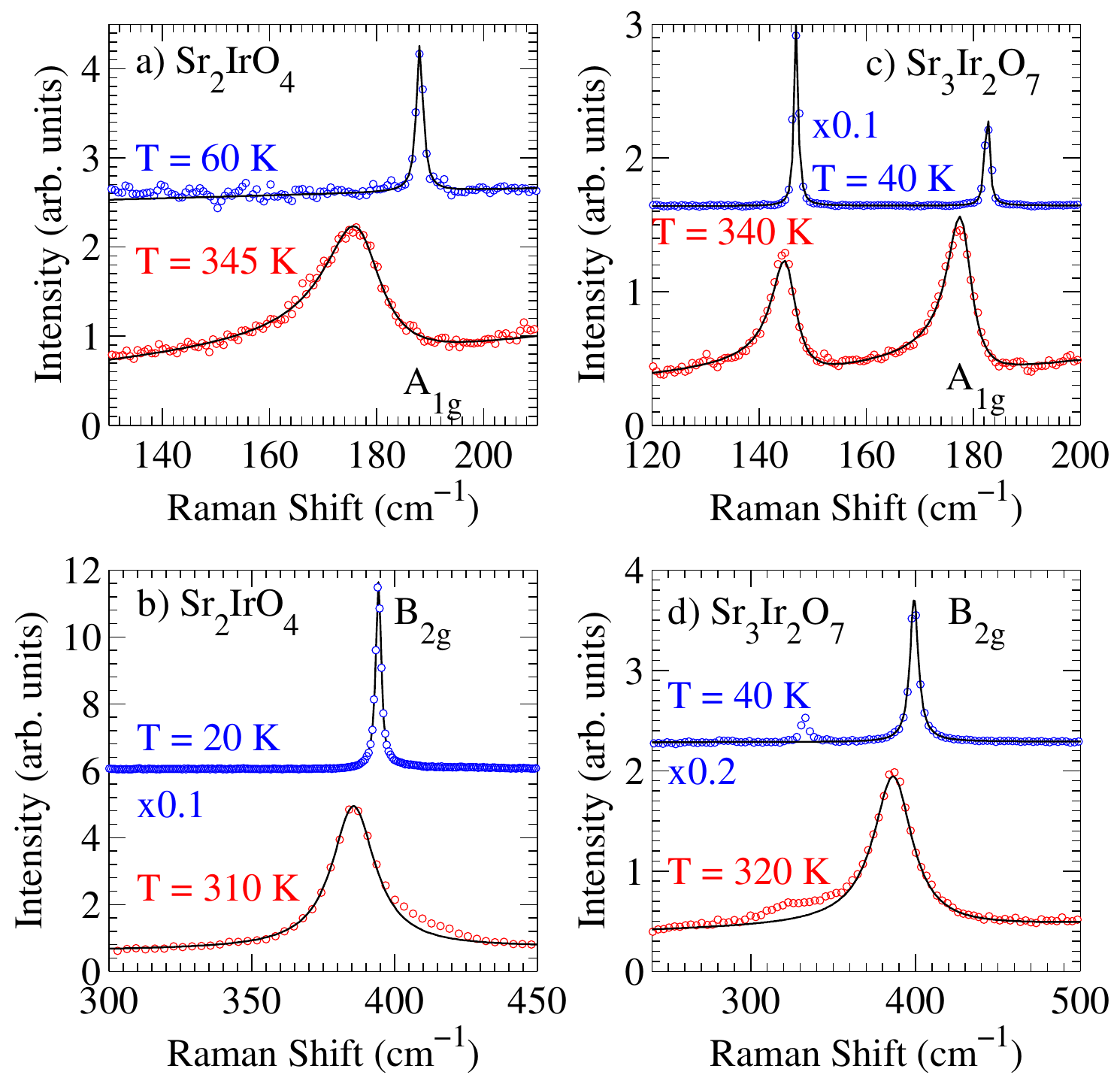}
\caption{\label{PhononsTemperture} Raman spectra of the (a) $A_{1g}$ and (b) $B_{2g}$ phonon modes of \SIO\ taken at high- and low-temperatures. In (c) and (d) the equivalent phonons for \bSIO\ can be seen.  The solid black lines are a fit to a Fano line shape. Low-temperature spectrum have been shifted and scaled in intensity (where indicated) for clarity. }
\end{figure}

Motivated by these findings, we have carefully investigated the Raman-active phonons in \SIO\ and \bSIO, whose lineshapes are sensitive indicators of coupling to spin and charge excitations. The Raman selection rules in \SIO\ have been discussed in Refs. \onlinecite{Sr214_Raman,Sr214_Ru_Raman}. Our focus is on the lineshapes and their temperature dependence, which are highlighted in Fig. \ref{PhononsTemperture} for two representative phonons: an intense $B_{2g}$ mode seen close to 380 cm$^{-1}$ in both \SIO\ and \bSIO, and a set of lower-energy $A_{1g}$ modes (one mode at 176 cm$^{-1}$ in \SIO, two at 140 and 170 cm$^{-1}$ in \bSIO). As shown in Fig. \ref{PhononsTemperture}(a) and (c), the latter modes display a pronounced Fano asymmetry at high-temperature ($T>T_N$) which vanishes for $T<T_N$.
The $B_{2g}$ mode of \SIO (\ref{PhononsTemperture}(b)) displays a more symmetric lineshape at high-temperature. (Note that there is some additional spectral weight on its high-energy side which is present also at low temperature and cannot be fitted properly using a Fano lineshape.) In \bSIO\ , however, it also exhibits a slight asymmetry.

To quantify the observed changes in Figs. \ref{PhononsTemperture} (a),(c), and (d) we have fitted the temperature evolution of the spectrum with Fano profiles  described by the formula: $I(\omega)=I_0(q+\epsilon)^2/(1+\epsilon^2)$, where $\epsilon=(\omega-\omega_0)/(\Gamma)$, $\omega_0$ is the bare phonon frequency, $\Gamma$ is the linewidth, and $q$ is the asymmetry parameter \cite{Fano:PR1961}. The $B_{2g}$ mode in panel (b) was fitted with a Lorentzian function. In Fig. \ref{IntMagnon} we plot the resulting frequency shift, $\omega_0$, and the Fano asymmetry parameter, $q$, of each phonon as a function of temperature. When fitting the two $A_{1g}$ modes in \bSIO\ the same values for $\Gamma$ and  $q$ could be used. Since these two modes behave in a very similar manner, we only focus on the higher-energy $A_{1g}$ mode. The dashed vertical line corresponds to the onset of magnetic order in \SIO\ and \bSIO, $T_N=240$ K and $T_N=285$ K respectively.


We first discuss the temperature dependence of the phonon frequencies which generally harden upon cooling, as expected from anharmonic phonon-phonon interactions. By fitting the high-temperature (200 K $\lesssim$ T) phonon frequency shift to standard anharmonic-decay model (solid lines in Fig. \ref{IntMagnon}) \cite{AnharmonicDecay} it becomes clear that some of these modes exhibit anomalous temperature dependence. In particular, the $B_{2g}$ (Fig. \ref{IntMagnon}(e)) and the $A_{1g}$ modes (Fig. \ref{IntMagnon}(d)) in \bSIO\ harden anomalously for T $\lesssim$ 200 K while the (a) $A_{1g}$ modes in \SIO\ soften anomalously just below the N\'eel temperature. Magnetic-order-induced phonon frequency anomalies of similar magnitude have been observed in other Mott insulators such as LaMnO$_3$ \cite{Podobedov_LMO3_Raman:PRB1998,Granado} and are usually attributed to the dependence of the superexchange interaction on the atomic coordinates, which are dynamically modulated in a lattice vibration.

In contrast, the observation of Fano lineshapes is highly unusual for phonons in Mott insulators. The pronounced Fano asymmetry observed for the $A_{1g}$ modes above $T_N$ (Fig. \ref{PhononsTemperture}(a) and (c)) indicates strong coupling to a continuum of excitations. The onset of magnetic order quenches this coupling, so that the Fano asymmetry parameter $q \rightarrow 0$ for $T < T_N$ (Fig. \ref{IntMagnon} (a) and (d)). In principle, such a continuum could arise from thermally excited charge carriers that are coupled to phonons in a manner analogous to correlated-electron metals where Fano lineshapes are widely observed in Raman scattering. Indeed, substantial changes of both the $dc$ and the optical conductivity have been reported for $T \sim T_N$ in \bSIO, where the Mott gap is comparatively small \cite{Cao_TransportSr327:PRB2002,Park_TdepOptics:PRB2014}. In \SIO, however, the Mott gap is much larger, \cite{Moon_TdepOptics:PRB2009} and the charge transport properties are virtually unaffected by the onset of magnetic order. \cite{Sr2IrO4_Transport}. Since the magnetic-order-induced phonon lineshape renormalization is of similar magnitude in both compounds (Fig. \ref{IntMagnon}(a) and (d)), charge excitations can be ruled out as the origin of these anomalies.

These considerations imply that the electronic continuum in the paramagnetic state is formed by the pseudospin degrees of freedom that order below $T_N$. The large electron-phonon linewidths manifest strong interactions of this continuum with lattice vibrations. This interaction can arise from partial admixture of the \jeff $= \frac{3}{2}$ level into the \jeff $= \frac{1}{2}$ ground state induced by crystalline electric fields, which actuates low-energy shape fluctuations of the valence-electron cloud that couple effectively to phonons. Electronic structure calculations for \SIO\ indeed indicate substantial mixing of \jeff $= \frac{1}{2}$ and $\frac{3}{2}$ states, supporting this scenario. \cite{Watanabe} As the pseudospin excitations condense into sharp, dispersive magnon and spin-orbit exciton modes below $T_N$, these fluctuations are quenched, and the phonon anomalies disappear. The strong broadening of the two-magnon peak with increasing temperature (Fig. \ref{IntMagnon}(c)) may then be understood as a converse manifestation of this pseudospin-lattice interaction. A related example was found in Ca$_2$RuO$_4$, a compound with $4d$ valence electrons and consequently smaller spin-orbit coupling that undergoes a sequence of correlation-driven metal-insulator and antiferromagnetic phase transitions upon cooling \cite{Cooper_Ca214:PRB2005}. Fano lineshapes of phonons in the temperature range between both transitions were attributed to orbital fluctuations. In contrast, phonons in orbitally non-degenerate Mott-insulating cuprates with $S = \frac{1}{2}$ electrons exhibit neither Fano lineshapes nor magnetic-order-induced phonon anomalies.


In summary, our Raman scattering study of stoichiometric, Mott-insulating \SIO\ and \bSIO\ has uncovered evidence of strong pseudospin-lattice interactions, which indicate that the spin-orbit coupling is not sufficiently strong to quench the orbital fluctuations in the paramagnetic state. The quantitative description of this interaction and an assessment of its consequences for the electronic properties of doped spin-orbit Mott-insulators are interesting challenges for future theoretical work.

\acknowledgements{We would like to thank G. Jackeli, G. Khaliullin, F. Mila, H. Takagi, S. Weidinger, and W. Zwerger for fruitful discussions. We acknowledge financial support by the DFG under grant TRR80.}


\end{document}